\documentclass [prc,aps,showpacs,twocolumn]{revtex4}
\usepackage{graphicx}

\begin{document}
% \oddsidemargin 5mm
% \baselineskip=20pt
% \large{
% \draft

\title{Equilibrium statistical mechanics for incomplete nonextensive statistics}

\author{A.S.~Parvan$^{a,b}$  and  T.S.~Bir\'{o}$^{c,}$}

\affiliation{$^{a}$Bogoliubov Laboratory of Theoretical Physics, Joint Institute for Nuclear Research, 141980 Dubna, Russian Federation}

\affiliation{$^{b}$Institute of Applied Physics, Moldova Academy of Sciences, MD-2028 Chisinau, Republic of Moldova}

\affiliation{$^{c}$KFKI Research Institute for Particle and Nuclear Physics, H-1525 Budapest, P.O.Box 49, Hungary}

%\email{parvan@theor.jinr.ru}
%\thanks{(Alexandru Parvan)}

%\date{\today}

\begin{abstract}
The incomplete nonextensive statistics in the canonical and microcanonical ensembles is explored in the general case and in a particular case for the ideal gas. By exact analytical results for the ideal gas it is shown that taking the thermodynamic limit, with $z=q/(1-q)$ being an extensive variable
of state, the incomplete nonextensive statistics satisfies the requirements of equilibrium thermodynamics. The thermodynamical potential of the statistical ensemble is a homogeneous function of the first degree of the extensive variables of state. In this case, the incomplete nonextensive statistics is equivalent to the usual Tsallis statistics. If $z$ is an intensive variable of state, i.e. the entropic index $q$ is a universal constant, the requirements of the equilibrium thermodynamics are violated.
\end{abstract}


\maketitle
% \narrowtext

%%%%%%%%%%%%%%%%%%%%%%%%%%%%%%%% SECTION 1 %%%%%%%%%%%%%%%%%%%%%%%%
\section{Introduction}
In modern physics there exist alternative theories for the equilibrium statistical mechanics: the Boltzmann-Gibbs statistics~\cite{Gibbs,Balescu}, the Tsallis statistics~\cite{Tsal88,Tsal98}, the R\'enyi statistics~\cite{Lenzi00,ParvBiro} based on R\'enyi entropy~\cite{Renyi,Wehrl}, the incomplete nonextensive statistics~\cite{Wang01_1,Wang02_1}. All of them are obtained from the usual Boltzmann-Gibbs statistics due to the modification of
the statistical entropy formula, the norm equation for a distribution function, and/or the way of taking ensemble averages. These statistics are defined at the equilibrium by the Jaynes principle of maximum entropy~\cite{Jaynes}. Therefore, they have to satisfy all requirements of equilibrium thermodynamics and, moreover, must be in agreement with probability theory and general principles of physics. The incomplete nonextensive statistics is defined by the Tsallis entropy formula~\cite{Tsal88}, the modified norm equation for a phase distribution function and a modified expectation value
for dynamical variables~\cite{Wang01_1,Wang02_1}. It differs from the Tsallis statistics only in the definition of the norm equation and ensemble averages. Nevertheless, there exist some other variants of incomplete nonextensive statistics which are defined either by an entropy different from the Tsallis one~\cite{Wang03_1,Wang04_1,Lima04,Sattin05} or by incomplete expectation values, but with the usual norm equation for a distribution function~\cite{Wang03_2}.

The main aim of this Letter is to investigate the relation of the original incomplete nonextensive statistics~\cite{Wang01_1,Wang02_1} to equilibrium thermodynamics, and to compare this with the one valid for the Tsallis statistics. A necessary and sufficient condition for fulfilling the zeroth law of thermodynamics, the principle of additivity, the Euler theorem, and the Gibbs-Duhem relation is that  the thermodynamic potential of the statistical ensemble be a homogeneous function of the first degree of the extensive variables of state~\cite{Parv1,Parv2}. This is achieved in the thermodynamic limit. For finite systems the thermodynamical potential is an inhomogeneous function. The correct definition of the thermodynamic limit for the Tsallis nonextensive statistics was first considered in Botet et al.~\cite{Botet1,Botet2} and in~\cite{Parv1,Parv2,ParvBiro06}. The problem of the zeroth law of thermodynamics for the incomplete nonextensive statistics was discussed in Refs.~\cite{Wang04_2,Wang06_1,Wang02_2,Wang02_3,Wang02_4,Wang02_5,Wang05_1,Wang05_2,Wang08_1,ParvBiro06}.

The main relations for the incomplete nonextensive statistics in finite systems for the canonical and microcanonical ensembles were derived in our previous paper~\cite{ParvBiro}. In the present Letter we discuss whether the incomplete nonextensive statistics can be thermodynamically consistent, i.e., whether the thermodynamical potential of the statistical ensemble is a homogeneous function of the first degree of its extensive variables of state. We concentrate on the thermodynamic limit, with $z$ being an extensive variable of state.

This Letter is organized as follows. In Section 2 we briefly outline the microcanonical ensemble for the incomplete nonextensive statistics in general and in particular for the ideal gas, and check the zeroth law of thermodynamics. The same procedure for the canonical ensemble is given in Section 3.

%%%%%%%%%%%%%%%%%%%%%%%%%%%%%%%% SECTION 2 %%%%%%%%%%%%%%%%%%%%%%%%
\section{Microcanonical ensemble}
\subsection{General formalism $(E,V,z,N)$}
The equilibrium statistical mechanics is an incomplete nonextensive statistics if the equilibrium phase space distribution function satisfies conditions imposed by the Tsallis' statistical entropy~\cite{Tsal88},
\begin{eqnarray}\label{1} \nonumber
S &=& -k \int d\Gamma \frac{\varrho-\varrho^{q}}{1-q} = \\
&=& k(z+1) \int d\Gamma \varrho^{\frac{z}{z+1}}(1-\varrho^{\frac{1}{z+1}}),
\end{eqnarray}
Here $\varrho$ satisfies an incomplete norm equation, and modified expectation values are taken as $A$~\cite{Wang01_1,Wang02_1},
\begin{eqnarray}\label{2}
1 &=& \int d\Gamma \varrho^{q} = \int d\Gamma \varrho^{\frac{z}{z+1}} , \\ \label{3}
 \langle A\rangle &=& \int d\Gamma \varrho^{q} A =\int d\Gamma \varrho^{\frac{z}{z+1}} A.
\end{eqnarray}
Here $f=\varrho^{q}=\varrho^{\frac{z}{z+1}}$ is the phase distribution function, $k$ is the Boltzmann constant, $q\in\mathbf{R}$ is a real parameter, $q\in [0,\infty]$, and $z$ is the variable of state, $z=q/(1-q)$.

Consider the microcanonical ensemble $(E,V,z,N)$. Using the thermodynamical method based on the fundamental equation of thermodynamics~\cite{Parv1} instead of the Jaynes principle~\cite{Jaynes}, we obtain the equilibrium phase distribution function and the statistical weight~\cite{ParvBiro}
\begin{eqnarray} \label{9}
 f &=& \varrho^{\frac{z}{z+1}}= W^{-1}\Delta(H-E)=\varrho_{G}, \\ \label{10}
 W &=& \int\limits_{D} d\Gamma=  \int\Delta(H-E) d\Gamma,
\end{eqnarray}
where $H$ is the Hamiltonian function, $\varrho_{G}$ is the distribution function of the Gibbs statistics, $z$ is the thermodynamical variable of state~\cite{Parv1} and $\Delta(\varepsilon)$ is the function distinct from zero only in the interval $0\leq\varepsilon\leq\Delta E$, where it is equal to unity, $D$ being the region of phase space restricted by inequality $E\leq H\leq E +\Delta E$. These two methods provide us with the same results for the distribution function(see Ref.~\cite{Parv1} and Refs.~\cite{ParvBiro06,Wang09_1,Wang09_2}). In the following the subscript $G$ indicates the Gibbs statistics. The thermodynamical potential of the microcanonical ensemble $(E,V,z,N)$, the entropy $S$, has the form~\cite{ParvBiro}
\begin{equation}\label{11}
     S = k(z+1) \left[1-e^{-\frac{S_{G}}{k z}}\right], \qquad  S_{G} = k\ln W,
\end{equation}
where $S_{G}$ is the usual Boltzmann-Gibbs entropy in the microcanonical ensemble. Since the distribution function (\ref{9}) and the entropy (\ref{11}) have equilibrium values, we obtain the fundamental equation of thermodynamics
\begin{equation}\label{13}
    TdS= dE + p dV + Xdz - \mu dN, \quad  \langle H\rangle=E,
\end{equation}
where
\begin{eqnarray}\label{14}
  \frac{1}{T} &=& \left(\frac{\partial S}{\partial E}\right)_{V,z,N}= \frac{z+1}{z T_{G}} \ e^{-\frac{S_{G}}{k z}}, \\ \label{16}
  X &=& T\left(\frac{\partial S}{\partial z}\right)_{E,V,N} =
   \frac{TS}{z+1}-\frac{T_{G}S_{G}}{z}, \;\;\;\; \\  \label{15}
  p &=& T\left(\frac{\partial S}{\partial V}\right)_{E,z,N}= p_{G},  \\ \label{17}
  \mu &=& -T\left(\frac{\partial S}{\partial N}\right)_{E,V,z}= \mu_{G},
\end{eqnarray}
where the definitions for the Gibbs measures can be found in Ref.~\cite{ParvBiro1}. The fundamental equation of thermodynamics (\ref{13}) provides the first and second principles of equilibrium thermodynamics
\begin{equation}\label{18}
    \delta Q = TdS, \quad  \delta Q= d\langle H\rangle + pdV + Xdz -\mu dN,
\end{equation}
where $\delta Q\ge 0$ is a heat transfer by the system to the environment during a quasistatic transition of the system from one equilibrium state to a nearby one.

The zeroth law of thermodynamics, the Euler theorem and the Gibbs-Duhem relation for the incomplete nonextensive statistics in the microcanonical ensemble are valid whenever the thermodynamic potential $S$ is a homogeneous function of the first degree of {\em all} extensive variables of state~\cite{Parv1}. For the homogeneous entropy $S$ all functions of variables of state are either homogeneous functions of the first degree (extensive or additive of the first degree $\mathcal{A}=\mathcal{A}_{1}+\mathcal{A}_{2}$)
\begin{eqnarray}\label{25}
  \mathcal{A}(E,\lambda V,\lambda z,\lambda N) &=& \lambda \mathcal{A}(E,V,z,N) \mbox{(z-extensive)}, \;\;\;\;\;\;\;  \\ \label{26}
  \mathcal{A}(E,\lambda V,z,\lambda N)  &=& \lambda \mathcal{A}(E,V,z,N)  \mbox{(z-intensive)}
\end{eqnarray}
or homogeneous functions of the zero degree (intensive or additive of the zero degree $\phi=\phi_{1}=\phi_{2}$)
\begin{eqnarray}\label{10bb}
  \phi(E,\lambda V,\lambda z,\lambda N) &=& \phi(E,V,z,N)  \mbox{(z-extensive)}, \;\;\;\;  \\ \label{11bb}
  \phi(E,\lambda V,z,\lambda N)  &=& \phi(E,V,z,N)  \mbox{(z-intensive)}.
\end{eqnarray}
Here $\lambda$ is an appropriate scaling parameter; it can be $\lambda=1/E$, $\lambda=1/N$, $\lambda=1/V$ ( or $\lambda=1/z$ for $z$ extensive). Note that the extensive function $\mathcal{A}$ is additive, $\mathcal{A}=\mathcal{A}_{1}+\mathcal{A}_{2}$, but the function $\lambda \mathcal{A}$ is intensive, $\lambda \mathcal{A}=\lambda_{1} \mathcal{A}_{1}=\lambda_{2} \mathcal{A}_{2}$. The variables of state should satisfy the following conditions in equilibrium:
\begin{eqnarray}\label{20}
  E &=& E_{1}+E_{2},  N = N_{1}+N_{2},  V=V_{1}+V_{2},  \\ \label{21}
 z &=& z_{1}+z_{2}  \qquad \mbox{(z-extensive)},  \\ \label{22}
z &=& z_{1}=z_{2} \qquad \mbox{(z-intensive)}.
\end{eqnarray}
For extensive variables $E,V,z,N$ the specific ratios $\lambda E$, $\lambda V$, $\lambda N$ and $\lambda z$ are intensives. If the entropy $S$ is a homogeneous function of the first degree of all extensive variables of state (\ref{25}), (\ref{26}), then for the intensive functions in the microcanonical ensemble (\ref{14}), (\ref{15}), (\ref{17}), we have
\begin{eqnarray}\label{7x}
  T &=& T_{1} = T_{2}, \quad p = p_{1} =  p_{2}, \quad \mu = \mu_{1} = \mu_{2}. \;\;\;\;
\end{eqnarray}
Therefore, the zeroth law of thermodynamics, $T = T_{1}=T_{2}$, is satisfied whenever the thermodynamic potential $S$ of the microcanonical ensemble is a homogeneous function of the first degree of the extensive variables of state~\cite{Parv1}. In this case, the entropy, its differential and the heat transfer are additive functions $S=S_{1}+S_{2}$, $dS = dS_{1}+ dS_{2}$ and $\delta Q = \delta Q_{1} + \delta Q_{2}$, respectively. The last equation follows from eq. (\ref{18}). The homogeneous $S$ satisfies the Euler theorem and the Gibbs-Duhem relation
\begin{eqnarray} \label{31}
 T S &=& \langle H\rangle+p V +Xz -\mu N  \quad \mbox{(z-extensive)}, \;\;\; \\ \label{31a}
  SdT &=& Vdp +zdX -Nd\mu
\end{eqnarray}
and
\begin{eqnarray} \label{32}
 T S &=& \langle H\rangle+p V -\mu N  \quad \mbox{(z-intensive)}, \;\; \\  \label{32a}
SdT &=& Vdp -Xdz -Nd\mu.
\end{eqnarray}
For $z$ being extensive, the Euler theorem  (\ref{31}) contains the term $Xz$ and it is consistent with the fundamental equation of thermodynamics (\ref{13}) as well as with the Gibbs-Duhem relation (\ref{31a}). However, for $z$ being intensive the Gibbs-Duhem relation (\ref{32a}) is violated, and the Euler theorem (\ref{32}) is not consistent  with the fundamental equation of thermodynamics (\ref{13}). Thus, in this case the variable of state $z$ can only be extensive.

The Tsallis entropy (\ref{1}), (\ref{11}) may still become a homogeneous function in the thermodynamic limit. For $z$ being {\it extensive} in this limit, functions of the variables of state are expanded in power series under the conditions $N\rightarrow\infty$, $E\rightarrow\infty$, $V\rightarrow\infty$, $z\rightarrow\infty$ and $\varepsilon=E/N =\mathrm{const}$, $v=V/N =\mathrm{const}$, $\tilde{z}=z/N =\mathrm{const}$ $(\alpha> 0)$. One determines
\begin{eqnarray} \label{35}
   \mathcal{A}(E,V,z,N) &=& N[a(\varepsilon,v,\tilde{z}) + O(N^{-\alpha})], \\ \label{37}
\phi(E,V,z,N) &=& \phi(\varepsilon,v,\tilde{z}) + O(N^{-\alpha}),
\end{eqnarray}
where $\varepsilon=E/N$ is the specific energy, $v=V/N$ is the specific volume, $\tilde{z}=z/N$ is the specific $z$, and $a =\mathcal{A} /N$ is the specific $\mathcal{A}$. For  $z$ being {\it intensive}, functions of the variables of state are expanded in power series under similar conditions as before, but with $z=\mathrm{const}$.
%the conditions  $N\rightarrow\infty$, $E\rightarrow\infty$, $V\rightarrow\infty$,  and $\varepsilon=E/N =\mathrm{const}$, $v=V/N =\mathrm{const}$, $z =\mathrm{const}$ $(\alpha> 0)$
\begin{eqnarray} \label{36}
   \mathcal{A}(E,V,z,N) &=& N[a(\varepsilon,v,z) + O(N^{-\alpha})], \\ \label{38}
\phi(E,V,z,N) &=& \phi(\varepsilon,v,z) + O(N^{-\alpha}).
\end{eqnarray}
One concludes that the entropy is a homogeneous function of the first degree in the thermodynamic limit.

\subsection{Ideal gas in microcanonical ensemble}
Let us investigate the thermodynamical properties of the microcanonical ideal Maxwell-Boltzmann gas~\cite{ParvBiro} in terms of the variable $z=q/(1-q)$. The phase distribution function (\ref{9}) and the statistical weight (\ref{10}) for the ideal gas in the microcanonical ensemble can be found in Ref.~\cite{ParvBiro1}. The thermodynamic potential (\ref{11}) is found to be
\begin{equation}\label{41}
    S=k (z+1) \left(1- W^{-1/z}\right),
\end{equation}
where $S_{G} = k \ln W$ is the Gibbs entropy. The temperature (\ref{14}), the pressure (\ref{15}), the variable $X$ (\ref{16}) and the chemical potential (\ref{17}) are now
\begin{eqnarray}\label{42}
  T &=& T_{G}\frac{z W^{1/z}}{z+1}, \qquad  p = p_{G}, \\  \label{44}
  X &=& kT_{G} \left[ \frac{z (W^{1/z}-1)}{z+1} -\frac{\ln W}{z}  \right],  \mu = \mu_{G},
\end{eqnarray}
where the definitions for the Gibbs measures can be found in Ref.~\cite{ParvBiro1}. The heat capacity is given by
\begin{equation}\label{47}
    C_{E} = \frac{k (z+1)}{W^{1/z}} \left(1+\frac{kzT_{G}}{E}\right)^{-1}.
\end{equation}
Its general microcanonical definition can be found in Ref.~\cite{ParvBiro1}. In the Gibbs limit the entropy (\ref{41}), the temperature and the variable $X$ can be written as  $S|_{z\to\pm\infty} = S_{G}$, $T|_{z\to\pm\infty} = T_{G}$ and $X|_{z\to\pm\infty}=0$, respectively.

Let us investigate the thermodynamics of the ideal gas in a finite system. We divide the system into two parts, under the conditions (\ref{20})--(\ref{22}). Then from Eqs.~(\ref{41})--(\ref{42}) it follows that the Tsallis entropy for a finite system is nonadditive (nonextensive) function, $S\neq S_{1}+S_{2}$, and its proper temperature is nonintensive function, $T\ne T_{1} \ne T_{2}$. Accordingly, the microcanonical phase distribution function for this statistics does not factorize, $\varrho \neq \varrho_{1}\varrho_{2}$, where $W \neq W_{1}W_{2}$. Using Eqs.~(\ref{41})--(\ref{44}), we obtain the Euler theorem for the incomplete nonextensive statistics in the case of $z$ being extensive
\begin{eqnarray}\label{52}
  TS + O_{W}^{(e)} &=&  E +pV +Xz -\mu N, \\ \label{53}
  O_{W}^{(e)} &=& O_{G} - kT_{G}\frac{z}{z+1}(W^{1/z}-1)
\end{eqnarray}
and in the case of $z$ being intensive
\begin{eqnarray}\label{54}
  TS + O_{W}^{(i)} &=&  E +pV -\mu N, \\ \label{55}
  O_{W}^{(i)} &=& O_{G} - kT_{G}z[W^{1/z}-1-\frac{1}{z}\ln W], \;\;\;\;\;\;
\end{eqnarray}
where
\begin{eqnarray}\label{51}\nonumber
   O_{G} &=& E + kT_{G}\ln\left[EN!\Gamma\left(\frac{3}{2}N\right)\right] + \\
  &+& kT_{G}N\left[1 - \psi(N+1)- \frac{3}{2}\psi\left(\frac{3}{2}N\right)\right]. \;\;\;\;\;\;\;
\end{eqnarray}
We conclude that for a finite ideal gas in the microcanonical ensemble, using the incomplete nonextensive statistics, the zeroth law of thermodynamics, the principle of statistical independence and the Euler theorem are violated. They are polluted by finite-size effects.

\subsection{Microcanonical ideal gas in the thermodynamic limit with $z$ being intensive}
Let us now consider the ideal gas in the thermodynamic limit with $z$ being intensive. The phase distribution function $\varrho$ and the statistical weight $W$ in the thermodynamic limit can be found in Ref.~\cite{ParvBiro1}. It is readily seen that the entropy (\ref{41}), the temperature and pressure (\ref{42}), the variable $X$ and the chemical potential (\ref{44}) can be written as
\begin{equation}\label{59}
  S = k (z+1) \left(1- w^{-N/z}\right)
\end{equation}
and
\begin{eqnarray}  \label{60}
  T &=& T_{G}\frac{z w^{N/z}}{z+1}, \quad  p = p_{G}, \\  \label{62}
  X &=& kT_{G}\left[\frac{z(w^{N/z}-1)}{z+1} -\frac{N \ln w }{z} \right], \;  \mu = \mu_{G}, \;\;\;\;
\end{eqnarray}
where the Gibbs measures in the thermodynamic limit can be found in Ref.~\cite{ParvBiro1}. Finally, the heat capacity (\ref{47}) is found to be $ C_{E} = k (z+1) w^{-N/z} \left(1+\frac{2z}{3N}\right)^{-1}$. In the Gibbs limit the incomplete nonextensive statistics quantities (\ref{59})--(\ref{62}) recover their Gibbs values, the entropy $S|_{z\to\pm\infty} = S_{G}$, the temperature $T|_{z\to\pm\infty} = T_{G}$, the variable $X|_{z\to\pm\infty}=0$ and the heat capacity $C_{E}|_{z\to\pm\infty} = C_{E,G}=3Nk/2$, respectively.
{\em Therefore, for $z$ being intensive in the microcanonical ensemble the Gibbs limit is commutative with the thermodynamic limit.}

Considering Eqs.~(\ref{20}), (\ref{22}), $\varepsilon=\varepsilon_{1}=\varepsilon_{2}$ and $v=v_{1}=v_{2}$,
in the thermodynamic limit for $z$ being intensive, we obtain that the entropy (\ref{59}) is
nonadditive. The temperature (\ref{60}) is nonintensive and the phase distribution function factorizes, $\varrho = \varrho_{1} \varrho_{2}$. The Euler theorem (\ref{54}) for the incomplete nonextensive statistics is violated
\begin{eqnarray} \label{69}
  O_{W}^{(i)} &=&  - kT_{G}z[w^{N/z}-1-\frac{N}{z}\ln w].
\end{eqnarray}
We conclude that the zeroth law of thermodynamics is not satisfied for the ideal gas in the thermodynamic limit with $z$ being intensive.

\subsection{Microcanonical ideal gas in the thermodynamic limit with $z$ being extensive}
Finally, we consider the thermodynamic limit with $z$ being extensive. The phase distribution function $\varrho$ and the statistical weight $W$ in the thermodynamic limit are calculated as in Ref.~\cite{ParvBiro1}. The entropy (\ref{41}), the temperature and pressure (\ref{42}), the variable $X$ and the chemical potential (\ref{44}) can be written as
\begin{equation}\label{70}
  S = k \tilde{z}N \left(1- w^{-1/\tilde{z}}\right)
\end{equation}
and
\begin{eqnarray}\label{71}
  T &=& T_{G} w^{1/\tilde{z}}, \qquad p = p_{G},  \\  \label{73}
  X &=& kT_{G} \left[w^{1/\tilde{z}}-1 -\frac{\ln w}{\tilde{z}}\right], \quad \mu = \mu_{G}.
\end{eqnarray}
The Gibbs quantities, labelled by the index $G$, are given in Ref.~\cite{ParvBiro1}.
The heat capacity (\ref{47}) takes the form $ C_{E} = k \tilde{z}N w^{-1/\tilde{z}} \left(1+\frac{2\tilde{z}}{3}\right)^{-1}$. In the Gibbs limit the incomplete nonextensive statistics quantities recover their Gibbs values, the entropy $S|_{\tilde{z}\to\pm\infty} = S_{G}$, the temperature $T|_{\tilde{z}\to\pm\infty} = T_{G}$, the variable $X|_{\tilde{z}\to\pm\infty}=0$ and the heat capacity $C_{E}|_{\tilde{z}\to\pm\infty} = C_{E,G}=3Nk/2$, respectively.
Therefore, for $z$ being extensive the Gibbs limit commutes with the thermodynamic limit.
Let us comment that the results (\ref{70})--(\ref{73}) for the incomplete nonextensive statistics
coincide with respective results for the Tsallis statistics given in Ref.~\cite{Parv1}.

Considering Eqs.~(\ref{20}), (\ref{21}) and $\varepsilon=\varepsilon_{1}=\varepsilon_{2}$, $v=v_{1}=v_{2}$, and $\tilde{z}=\tilde{z}_{1}=\tilde{z}_{2}$, in the thermodynamic limit for $z$ being extensive, we obtain that the entropy (\ref{70}) is additive, the variable $X$ (\ref{73}) and the temperature (\ref{71}) are intensive and the phase distribution function factorizes, $\varrho = \varrho_{1} \varrho_{2}$. Moreover, the Euler theorem (\ref{52}) is valid, $O_{W}^{(e)}=0$. Thus, for the incomplete nonextensive statistics in the thermodynamic limit with $z$ being extensive the entropy is a homogeneous function of the first degree and the zeroth law of thermodynamics, the Euler theorem and the principle of statistical independence are valid.

\section{Canonical ensemble}
\subsection{General formalism $(T,V,z,N)$}
Let us investigate the canonical ensemble $(T,V,z,N)$. To obtain the equilibrium phase distribution function, we use the thermodynamical method explored in~\cite{Parv1,Parv2} instead of the Jaynes principle~\cite{Jaynes}
\begin{eqnarray}\label{82}
f=\varrho^{\frac{z}{z+1}} &=& \left[1+\frac{z}{(z+1)^{2}}\frac{\Lambda-H}{kT}\right]^{z},
\end{eqnarray}
where $z=q/(1-q)$ and $\Lambda =\Lambda(T,V,z,N)$ is the norm function which is the solution of Eq.~(\ref{2}). For details, see Ref.~\cite{ParvBiro}. These two methods provide us with the same results for the distribution function (see Ref.~\cite{Parv2} and Refs.~\cite{ParvBiro06,Wang09_1,Wang09_2}). The expectation value $\langle A \rangle$ of a dynamical variable $A$ is determined by Eq.~(\ref{3}) with the distribution function (\ref{82}). The entropy and the free energy are expressed through the norm functions $\Lambda$ and $\langle H\rangle$ as
\begin{eqnarray}\label{85}
  S &=& \frac{z (\langle H\rangle -\Lambda)}{T (z+1)}, \\ \label{86}
  F & \equiv & \langle H\rangle -TS=\frac{\langle H\rangle+z \Lambda}{z+1}.
\end{eqnarray}
Equation (\ref{86}) is the Legendre transform of energy with respect to the entropy of the system. The fundamental equation of thermodynamics in the canonical ensemble is the same as Eq.~(\ref{13}) but with~\cite{ParvBiro}
\begin{eqnarray}\label{88}
  p &=& \int d\Gamma \varrho^{\frac{z}{z+1}} \left(-\frac{\partial H}{\partial V} \right)_{T,z,N}, \\ \label{89}
  \mu &=& \int  d\Gamma \varrho^{\frac{z}{z+1}} \left(\frac{\partial H}{\partial N} \right)_{T,V,z}, \\ \label{90}
  X &=& kT\int d\Gamma \varrho^{\frac{z}{z+1}} \nonumber \\
  &\times& \left[1-\varrho^{\frac{1}{z+1}} +\frac{z+1}{z} \varrho^{\frac{1}{z+1}} \ln\varrho^{\frac{1}{z+1}}\right],
\end{eqnarray}
where the following properties of the Hamiltonian function, $(\partial H/\partial T)_{V,z,N}=(\partial H/\partial z)_{T,V,N}=0$, were assumed.
It provides the first and the second laws of thermodynamics (\ref{18}). The thermodynamical relations for $S,p,X$ and $\mu$ in the canonical ensemble can be found in Ref.~\cite{Parv2}. Note that the thermodynamical relations for the quantum incomplete nonextensive statistics are the same as for the classical one given above.

The zeroth law of thermodynamics, the Euler theorem and the Gibbs-Duhem relation for the incomplete nonextensive statistics in the canonical ensemble are valid if the thermodynamic potential $F$ is a homogeneous function of the first degree of {\em all} extensive variables of state. The general proof of this statement is the same as the one given in Refs.~\cite{Parv2,ParvBiro1}. Moreover, the definition of the thermodynamical limit and the proof of the extensivity for the variable of state $z$ in the canonical ensemble are the same as for the R\'enyi statistics~\cite{ParvBiro1}.

\subsection{Ideal gas in the canonical ensemble}
Let us rewrite the main relations for the ideal gas in the canonical ensemble~\cite{ParvBiro} in terms of
the variable of state $z=q/(1-q)$. The phase space distribution function (\ref{82}) and the norm function $\Lambda$ can be written as
\begin{eqnarray}\label{115}
f &=& \varrho^{\frac{z}{z+1}}= \left[1+\frac{z}{(z+1)^{2}}\frac{\Lambda-\sum_{i=1}^{N}\frac{\vec{p}_{i}^{2}}{2m}}{kT}\right]^{z}, \\ \label{116}
\Lambda &=& kT \frac{(z+1)^{2}(B-1)}{z},  B = (a_{z,N}Z_{G})^{-\frac{1}{z+\frac{3}{2}N}} \;\;\;\;\;
\end{eqnarray}
and
\begin{eqnarray}\label{117}
  a_{z,N} &=& \frac{\Gamma(z+1)}
  {(\frac{z}{(z+1)^{2}})^{\frac{3}{2}N}\Gamma(z+1+\frac{3}{2}N)}, \quad  z>0, \\ \label{118}
  a_{z,N} &=& \frac{\Gamma(-z-\frac{3}{2}N)}
  {(\frac{-z}{(z+1)^{2}})^{\frac{3}{2}N}\Gamma(-z)}, \qquad  z<0,
\end{eqnarray}
where $g$ is the spin-isospin degeneracy factor and $z<-\frac{3}{2}N$ in Eq.~(\ref{118}). The partition function $Z_{G}$ for the ideal gas was defined in Ref.~\cite{Parv2}. The thermodynamic potential can be written as
\begin{eqnarray}\label{121}
    F &=& \frac{kT}{1+\frac{3N}{2(z+1)}} \left[\frac{3N}{2z} +
      \frac{z \Lambda}{(z+1)kT} (1+\frac{3N}{2z})\right]. \;\;\;\;\;
\end{eqnarray}
The energy, the entropy (\ref{85}) and the pressure (\ref{88}) can be written as
\begin{eqnarray}\label{122}
\langle H\rangle &=& \frac{3}{2} NkT \frac{z+1}{z} \frac{1+\frac{z}{(z+1)^{2}}\frac{\Lambda}{kT}}{1+\frac{3}{2}\frac{N}{z+1}}, \\ \label{123}
  S &=&  k \frac{\frac{3}{2}N-\frac{z}{z+1}\frac{\Lambda}{kT}}{1+\frac{3}{2}\frac{N}{z+1}}, \\ \label{124}
  p &=&  \frac{N}{V} kT \frac{z+1}{z} \frac{1+\frac{z}{(z+1)^{2}}\frac{\Lambda}{kT}}{1+\frac{3}{2}\frac{N}{z+1}}=
  \frac{2}{3} \frac{\langle H\rangle}{V}.
\end{eqnarray}
The chemical potential is found to be
\begin{eqnarray}\label{125}\nonumber
    \mu &=& \frac{\langle H\rangle}{N(z+1+\frac{3}{2}N)}- \\
   &-&\frac{\langle H\rangle}{N} \left[ \ln(1+\frac{z}{(z+1)^{2}}\frac{\Lambda}{kT}) +\frac{2}{3}(A_{\mu} - \frac{\mu_{G}}{kT}) \right], \;\;\;\;\;\;\;
\end{eqnarray}
where $A_{\mu}=\partial (\ln a_{z,N})/\partial N$ and
\begin{eqnarray}\label{126}
  \frac{2A_{\mu}}{3} &=& -\psi(z+1+\frac{3}{2}N)- \ln(\frac{z}{(z+1)^{2}}), z>0, \;\;\;\;\; \\ \label{127}
  \frac{2A_{\mu}}{3} &=& -\psi(-z-\frac{3}{2}N) - \ln(\frac{-z}{(z+1)^{2}}), z<0.
\end{eqnarray}
Here $\psi(z)$ is the psi-function, $z<-\frac{3}{2}N$ in Eq.~(\ref{127}). The chemical potential $\mu_{G}$ for the Gibbs statistics was given in Ref.~\cite{ParvBiro1}. The function $X$ can be written as
\begin{eqnarray}\label{129}\nonumber
    X &=& \frac{\langle H\rangle}{z(z+1)}\frac{1+\frac{3}{2}\frac{N}{(z+1)^{2}}}{1+\frac{3}{2}\frac{N}{z+1}}-
\frac{z \Lambda}{(z+1)^{2}}  + \\
 &+& \frac{2}{3}\frac{\langle H\rangle}{N} \left[ X_{a}+\ln\left(1+\frac{z}{(z+1)^{2}}\frac{\Lambda}{kT}\right) \right], \;\;\;\; \\
\label{130} \nonumber
  X_{a} &=& \frac{3N(z-1)}{2z(z+1)} - \\
  &-& \psi(z+1+\frac{3N}{2}) +  \psi(z+1), \quad z>0,  \\ \label{131} \nonumber
  X_{a} &=& \frac{3N(z-1)}{2z(z+1)} - \\
  &-& \psi(-z-\frac{3N}{2}) +  \psi(-z), \quad  z<-\frac{3}{2}N,
\end{eqnarray}
where $X_{a}=\partial (\ln a_{z,N})/\partial z$. The heat capacity is now $C_{VzN} =\langle H\rangle (1+\frac{3N}{2z})^{-1}/T$. Its definition in the canonical ensemble can be found in Ref.~\cite{ParvBiro1}. In the Gibbs limit $z\to\pm\infty,N=const,V=const$ the incomplete nonextensive statistics recovers the usual Boltzmann-Gibbs statistics. We conclude that for the finite ideal gas in the canonical ensemble the thermodynamical potential (\ref{121}) is an inhomogeneous function. Therefore, the zeroth law of thermodynamics, the Euler theorem and the principle of additivity are not satisfied.

\subsection{Canonical ideal gas in the thermodynamic limit with $z$ being extensive}
Let us consider the thermodynamical limit with $z$ being extensive, $N\to\infty$, $V\to\infty$, $z\to\pm\infty$ and $v=V/N=const$, $\tilde{z}=z/N=const$. The phase distribution function (\ref{115}) and the norm function $\Lambda$ (\ref{116}) take the form \begin{eqnarray}\label{141}
f &=& \varrho = \left[1+\frac{1}{\tilde{z}N} \frac{\Lambda-\sum_{i=1}^{N}
\frac{\vec{p}_{i}^{2}}{2m}}{kT} \right]^{\tilde{z}N},    \\ \label{142}
 \Lambda &=&  kT\tilde{z}N (B-1), B = (\tilde{Z}_{G}e^{3/2})^{-\frac{1}{\tilde{z}+\frac{3}{2}}} (1+\frac{3}{2\tilde{z}}), \;\;\;\;\;\; \\  \label{144}
 && a_{z,N} =  e^{\frac{3}{2}N}\left(1+\frac{3}{2\tilde{z}}\right)^{-(\tilde{z}+\frac{3}{2})N}.
\end{eqnarray}
The partition function $Z_{G}$, $\tilde{Z}_{G}$ can be found in Ref.~\cite{ParvBiro1}. The thermodynamic potential (\ref{121}) in the thermodynamic limit with $z$ being extensive takes the form
\begin{equation}\label{145}
    F  = kT\tilde{z}N (B-1) = \Lambda.
\end{equation}
Therefore, in the thermodynamic limit the norm function $\Lambda$ coincides with the free energy,
because in the series (\ref{145}) only the term proportional to the first power of $N$ is kept. The thermodynamical potential (\ref{145}) is a homogeneous function of first degree since $\tilde{z}$ is a intensive variable. The energy (\ref{122}), the entropy (\ref{123}) and the pressure (\ref{124}) in the thermodynamic limit can be written as
\begin{eqnarray}\label{147}
  S &=&  k N\tilde{z} \left(1- \frac{B}{1+\frac{3}{2\tilde{z}}}\right), \\ \label{146}
\langle H\rangle &=& \frac{3}{2} NkT \frac{B}{1+\frac{3}{2\tilde{z}}},
  p =  \frac{kT}{v} \frac{B}{1+\frac{3}{2\tilde{z}}} = \frac{2}{3} \frac{\langle H\rangle}{V}. \;\;\;\;\;
\end{eqnarray}
The energy (\ref{146}) and the entropy (\ref{147}) are extensive functions but the pressure $p$ is intensive. The variable of state $X$ (\ref{129}) and the chemical potential (\ref{125}) can be written as
\begin{eqnarray}\label{149}
    X &=& kT \left[1+\frac{B}{1+\frac{3}{2\tilde{z}}} \left(\ln\frac{B}{1+\frac{3}{2\tilde{z}}}-1\right) \right], \\ \label{150}
    \mu &=& -kT \frac{B}{1+\frac{3}{2\tilde{z}}} \left[\frac{3}{2} \ln\frac{B}{1+\frac{3}{2\tilde{z}}}
    -\frac{\mu_{G}}{kT} \right].
\end{eqnarray}
The chemical potential $\mu_{G}$ in the thermodynamic limit can be found in Ref.~\cite{ParvBiro1}. The heat capacity is now $ C_{VzN} =\langle H\rangle (1+\frac{3}{2\tilde{z}})^{-1}/T$. In the Gibbs limit $\tilde{z}\to\pm\infty$ and $N=const$ the incomplete nonextensive statistics recovers the usual Boltzmann-Gibbs statistics. For example, the measures (\ref{145}), (\ref{147}) take the form $F|_{\tilde{z}\to\pm\infty} = F_{G}$,  $S|_{\tilde{z}\to\pm\infty} = S_{G}$, respectively. This way in the incomplete nonextensive statistics, with $z$ being extensive, the Gibbs limit commutes with the thermodynamic limit. This conclusion differs from the results found in Ref.~\cite{Abe2}.

Dividing the system into two parts under the conditions $T=T_{1}=T_{2}$, $v=v_{1}=v_{2}$ and $\tilde{z}=\tilde{z}_{1}=\tilde{z}_{2}$, we obtain that the Tsallis entropy (\ref{147}) is a homogeneous function of first degree (extensive) and it is additive, $S = S_{1}+S_{2}$. Therefore, for the ideal gas in the canonical ensemble in the thermodynamic limit with $z$ extensive the zeroth law of thermodynamics, $T=T_{1}=T_{2}$, and the Euler theorem (\ref{31}) are valid. Let us note that in the thermodynamical limit with $z$ extensive all relations for the ideal gas in the canonical ensemble (\ref{141})--(\ref{150}) completely coincide with the ones of the original Tsallis statistics given in~\cite{Parv2}. Thus the incomplete nonextensive statistics satisfies all requirements of equilibrium thermodynamics in the canonical ensemble in the thermodynamical limit with $z$ extensive. The functions of the variables of state are homogeneous functions of first degree, for extensive, or homogeneous functions of the zero degree,for intensive variables. In particular, the temperature is an intensive variable and thus provides implementation of the zeroth law of thermodynamics.

\subsection{Canonical ideal gas in thermodynamical limit with $z$ being intensive}
We inspect an ideal gas in the thermodynamical limit with $z$ being intensive: $N\to\infty$, $V\to\infty$ and $v=V/N=const$, $z=const$.
For the sake of convenience let us consider only the case $z>0$. The phase space distribution function (\ref{115}) and the functions (\ref{116})--(\ref{117}) take the form
\begin{eqnarray}\label{158}
f=\varrho^{\frac{z}{z+1}} &=& \left[1+\frac{z}{(z+1)^{2}}\frac{\Lambda-\sum_{i=1}^{N}\frac{\vec{p}_{i}^{2}}{2m}}{kT}\right]^{z}, \\ \label{159}
 \Lambda &=&\frac{3}{2}N kT \tilde{Z}_{G}^{-2/3} e^{-1},
\end{eqnarray}
where
\begin{eqnarray} \label{160}
  a_{z,N} &=& \frac{\Gamma(z+1)  e^{z+1+\frac{3}{2}N}/\sqrt{2\pi} }
 {\left(\frac{z}{(z+1)^{2}}\right)^{\frac{3}{2}N} ( z+1+\frac{3}{2} N)^{z+\frac{1}{2}+\frac{3}{2}N}}. \;\;\;\;\;\;
\end{eqnarray}
The thermodynamic potential (\ref{121}) in the thermodynamic limit with $z$ being intensive takes the form $F=\Lambda$. Therefore, the norm function $\Lambda$ coincides with the free energy. In the series expansion of $F$ only the leading term proportional to the first power of $N$ is kept.
The energy (\ref{122}), the entropy (\ref{123}) and the pressure (\ref{124}) can be written as
\begin{eqnarray}\label{162}
\langle H\rangle &=& \Lambda, \qquad   S = \mathcal{O}(N^{0}), \\ \label{163a}
  \mathcal{O}(N^{0}) &=& k (z+1) \left(1- \frac{z}{z+1}\tilde{Z}_{G}^{-2/3} e^{-1}\right), \;\;\;\; \\ \label{164}
  p &=&  \frac{kT}{v} \tilde{Z}_{G}^{-2/3} e^{-1} = \frac{2}{3} \frac{\langle H\rangle}{V}.
\end{eqnarray}
Therefore, the entropy $S=0$. The function of state $X$ (\ref{129}) and the chemical potential (\ref{125}) can be written as $X=0$ and $\mu = \frac{5}{3} \frac{\Lambda}{N}$. The heat capacity is now $C_{VzN} =  \mathcal{O}(N^{0})$, where $\mathcal{O}(N^{0})=k \tilde{Z}_{G}^{-2/3} e^{-1}$.
In the Gibbs limit $z\to\pm\infty$ and $N=const$ the distribution function (\ref{158}) and the norm function (\ref{159}) do not resemble the distribution function and the free energy of the Gibbs statistics. The variables $F,\langle H\rangle,S,p,X,\mu$ and $C_{VzN}$ also do not resemble their Gibbs values. Thus, for the incomplete nonextensive statistics with $z$ being intensive the Gibbs limit does not commute with the thermodynamic limit as in Ref.~\cite{Abe2}. This means that the thermodynamic limit with $z$ being intensive is erroneous for both the incomplete nonextensive statistics and the Tsallis one~\cite{Abe2}.

The Euler theorem (\ref{32}) is valid if for extensive functions we keep the terms proportional to the first power of $N$ and for intensive ones we keep the terms proportional to $N^{0}$. In the thermodynamic limit with $z$ being intensive, the entropy $S$ is intensive and it is a homogeneous function of the zeroth degree. This contradicts the requirements of the equilibrium thermodynamics. Moreover, the thermodynamic limit with $z$ being intensive is not consistent with the Gibbs limit. Therefore, the ideal gas results with $z$ being intensive completely contradict the equilibrium thermodynamics.

%%%%%%%%%%%%%%%%%%%%%%%%%% CONCLUSION %%%%%%%%%%%%%%%%%%%%
\section{Conclusions}
In this Letter, we have examined the incomplete nonextensive statistics in the canonical and microcanonical ensembles in the general case and in the particular case of the classical ideal gas. The exact analytical results for the ideal gas were obtained for both finite systems and in the thermodynamic limit. The ideal gas in the thermodynamic limit was studied in two cases: the variable $z$ being extensive or intensive. Connections between incomplete nonextensive statistics and equilibrium thermodynamics were analyzed.

To this end, we summarize our main results. The phase distribution functions for the canonical and microcanonical ensembles were derived by the method based on the fundamental equation of thermodynamics. This method gives the same result as starting from the Jaynes principle. The connection of the incomplete nonextensive statistics with equilibrium thermodynamics was tested. The main thermodynamical relations were obtained from ensemble averages by using phase space distribution functions. It is shown that the zeroth law of thermodynamics, the principle of additivity, the Euler theorem and the Gibbs-Duhem relation are valid only if the corresponding thermodynamical potential is a homogeneous function of the first degree of all extensive variables of state. By analytical results for the ideal gas it was revealed that the thermodynamical potential can be a homogenous function of the first degree only in the thermodynamic limit, whenever the entropic index $z$ is an extensive variable of state. Moreover, it was found that the variable $z$ must be extensive in order the Euler theorem be consistent with the fundamental equation of thermodynamics and the Gibbs-Duhem relation.

On the whole, our analytical results for the ideal gas indicate that the incomplete nonextensive statistics in the canonical and microcanonical ensembles is thermodynamically consistent whenever it coincides with the usual Tsallis statistics in the thermodynamic limit with $z$ being an extensive variable of state. Therefore, we conclude that the incomplete nonextensive statistics in the equilibrium statistical mechanics duplicates the thermodynamic relations of the familiar Tsallis statistics and leads to no new result in this respect. This conclusion relies on the particular approach that the extra parameter $q$ in the entropy formula is treated as related to a {\em variable of state}, $z=q/(1-q)$.

%%%%%%%%%%%%%%%%%%%%%%%%%% ACKNOWLEDGMENT %%%%%%%%%%%%
{\bf Acknowledgments:} This research was partially supported by the joint research project of JINR and IFIN-HH, protocol N 3891-3-09/09, N 4006, the RFBR grant 08-02-01003-a and MTA-JINR grant, OTKA T49466. We acknowledge fruitful discussions with P.~Levai and V.D.~Toneev. The authors are also indebted to P\'eter V\'an for drawing their attention to Ref.~\cite{Renyi} on the first appearance of R\'eny entropy.

\end{document}